# Monolithic Silicon Carbide Metalenses

Otto Cranwell Schaeper[1,*], Ziwei Yang[2,*], Mehran Kianinia[1,4], Johannes E. Fröch[1], Andrei Komar[2], Zhao Mu[3,5], Weibo Gao[3,5], Dragomir Neshev[2], Igor Aharonovich[1,4]

[1]*School of Mathematical and Physical Sciences, Faculty of Science, University of Technology Sydney, Ultimo, New South Wales 2007, Australia*
[2]*ARC Centre of Excellence for Transformative Meta-Optical Systems (TMOS), Research School of Physics, The Australian National University, Canberra ACT 2600, Australia*
[3]*Division of Physics and Applied Physics, School of Physical and Mathematical Sciences, Nanyang Technological University, Singapore, 637371 Singapore*
[4]*ARC Centre of Excellence for Transformative Meta-Optical Systems (TMOS), University of Technology Sydney, Ultimo, New South Wales 2007, Australia*
[5]*The Photonics Institute and Centre for Disruptive Photonic Technologies, Nanyang Technological University, Singapore, 637371 Singapore*

#igor.aharonovich@uts.edu.au   ^dragomir.neshev@anu.edu.au
*equal contribution

**Abstract**
*Silicon carbide has emerged as a promising material platform for quantum photonics and nonlinear optics. These properties make the development of integrated photonic components in high-quality silicon carbide (SiC) a critical aspect for the advancement of scalable on-chip networks. In this work, we numerically design, fabricate and demonstrate the performance of monolithic metalenses from SiC suitable for on-chip optical operations. We engineer two distinct lenses, with parabolic and cubic phase profiles, operating in the near-infrared spectral range, which is of interest for quantum and photonic applications. We support the lens fabrication by optical transmission measurement and characterize the focal points of the lenses. Our results will accelerate the development of SiC nanophotonic devices and aid in an on-chip integration of quantum emitters with meta-optical components.*

Silicon Carbide (SiC) has attained increased attention and become a prime materials platform in nanophotonic and quantum photonic research over the last decade [1-14]. This status derives from the manifold attractive characteristics, including a low loss in the visible and infrared (IR) spectral range, availability of p- and n-type doping, a high $\chi^{(2)}$ nonlinear susceptibility, that are vital for a myriad of on-chip devices. In addition, the presence of several different optically active defects in the IR wavelength range with promising spin properties reignited the interest in silicon carbide as a platform for quantum photonics and quantum information processing [9, 11, 12].

In this context, the extraction efficiency of photons emitted from such defects becomes critical and is often limited due to the high refractive index of silicon carbide (n~2.6 at wavelength of 1 μm). Therefore, significant research has been devoted to the engineering of nanostructures from SiC, including photonic crystal cavities, ring resonators, and waveguides, to increase photon extraction [15-19]. While on-chip planar structures can aid some quantum photonic applications, they are typically restricted to thin films and limited in size. An alternative solution is the use of solid immersion lenses (SILs) that can be positioned on top, 3D printed or sculptured from the host material[20, 21].

Recently, solid immersion metalenses were suggested and realized in a diamond as a viable alternative to conventional SILs [22, 23]. The metalenses consist of periodically arranged nanostructures (e.g. cylinders, bars, etc), which scatter the incident light and impart a spatially varying phase delay onto the transmitted wave [24-27]. By tailoring this spatially distributed phase response, deliberate control over the transmitted light enables optical operations from focusing (metalenses), polarization control, beam steering, and uncountable other applications. For example, a collection lens from a point emitter would require a radial parabolic phase distribution [28], which is reciprocally equivalent to focusing of incoming beams inside the SiC material.

For large-scale photonic applications, including light-emitting diodes (LEDs) and single defect emitters, where silicon carbide is the frontrunner, metalenses are the best solution for collecting light from deep within the material. For example, for extraction of the light from a single emitter near the SiC surface, the use of high-numerical-aperture metalenses with unity efficiency [29] would be highly beneficial. Alternatively, for optically addressing and imaging of defects at an unknown depth in the material, metalenses with an extended focal length, incorporating cubic transverse phase profile [30-32] could further offer significant advantages. Overall, metalenses from silicon carbide can be used as independent devices in any application where lens behavior is required and the use of SiC material broadens the possible wavelength range. However, there is no demonstration of SiC metalenses, to date.

In this work, we design and demonstrate experimentally monolithic SiC metalenses, operating at the near-infrared spectral range, suitable for several photonic applications. The metalenses use phase alteration to elicit a response from the light that is positioned at the focal point of the lens - namely deep within the material. An additional advantage of the metalens is its immersion within the bulk material, which avoids the use of additional objective lenses for light collection. Specifically, we design and fabricate two types of metalenses - conventional parabolic-phase metalens and extended-focus metalens with a cubic transverse profile. We characterize the metalenses performance that shows good focusing in the near-infrared range. Our results constitute an important step for silicon carbide photonics, as it paves the way to a wider range of implementations and thus broadens the scope of potential applications for SiC photonics.

The immersed metalens unit cell is simulated by a finite difference time domain (FDTD) method (Lumerical), using a hexagonal lattice with periodic boundary conditions along *x*- and *y*-directions. A schematic of the metalens arrangement is shown in Figure 1a. In our simulations, the substrate and cylinder have the same anisotropic refractive index ($n_{xx} = n_{yy} = 2.58$, $n_{zz} = 2.63$). The working wavelength is 1,100 nm, which is chosen to match the emission wavelength of common defects in the SiC[11]. Due to the consistency of the refractive index of the substrate and the cylinder, the shift of the propagation phase is dominated by the waveguide and Fabry-Perot modes, whose transmittance and phase are shown in Figure 1b and c.

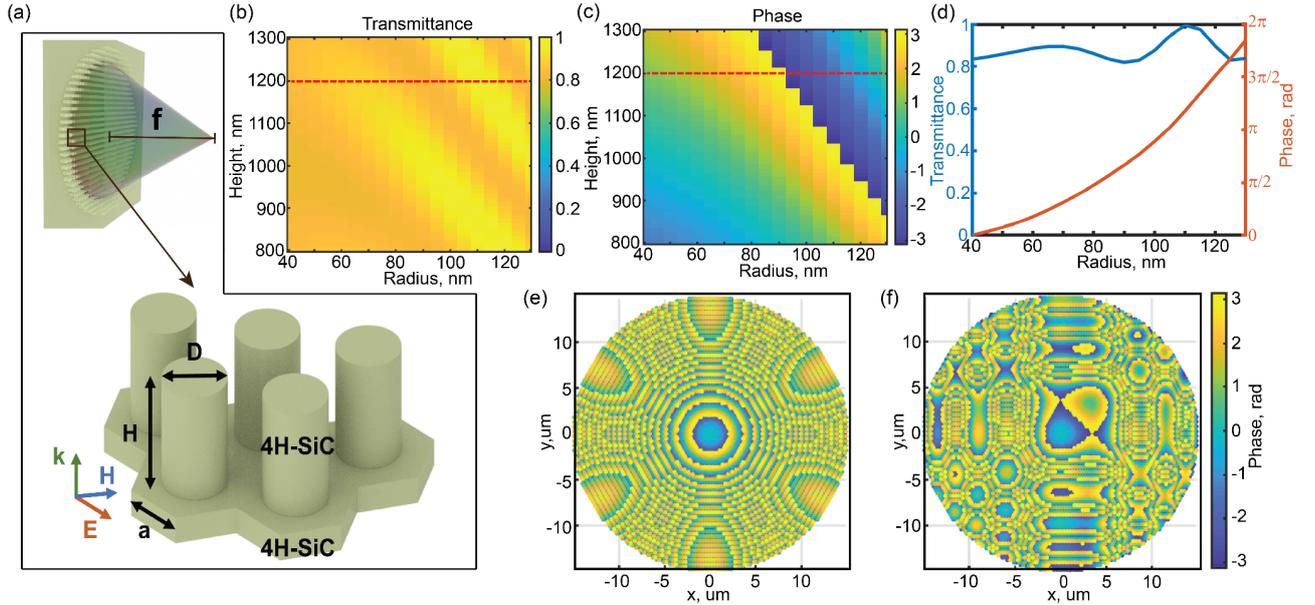

*Figure 1. Design of parabolic and cubic phase metasurfaces from monolithic silicon carbide. (a) Schematic illustration of the metalens and its components. D is the diameter of the pillar, H is the height of the pillar and a is lattice constant. (b, c) Simulated two-dimensional transmittance and phase for different cylinder radii and heights, respectively. (d) Transmittance and phase profiles of metasurfaces at a fixed height of H=1,200 nm, taken along the red dash line in (b, c). (e, f) Designed parabolic and cubic phase metalens, respectively, with a 30 μm diameter lens aperture.*

The results demonstrate that changing the radius and height of the individual cylinders allows obtaining the corresponding phase with high transmittance efficiency ( > 0.8). To achieve a 2π phase shift with a reliable radius range, the height of the cylinder has to be higher than the red dash line in Figure 1b, which is *H*=1,200 nm. We select the minimum height that can cover the 2π phase range with respect to the radius, as shown in Figure 1d, to relax the etching requirements in the nanofabrication process. The required phase can then be discretized from 0 to 2π based on different cylinder radius to constitute the total phase profile of the metalens. The utilized hexagonal lattice pattern arrangement in our metalens design allows for dense packing of the cylinders in the metalens and a better match to the circular shape of the overall lens. This hexagonal arrangement also leads to smoother phase sampling of each unit cell boundary zone, which yields a higher imaging quality than for a square lattice [24].

Utilizing the obtained phase to cylinder radius mapping, we designed two types of metalenses. The first is a conventional metalens with a parabolic phase profile and a converging wavefront. The phase profile of the first lens is shown in Figure 1e and can be described as

$$\varphi_p(x,y) = \frac{2\pi}{\lambda}\left(f - \sqrt{x^2 + y^2 + f^2}\right),$$

where λ is the incident light wavelength and $f$ is the focal length of the metalens. Every discrete phase point of the lens $\varphi_p$ corresponds to the transverse $x$ and $y$ position of the lens plane. Since the metalenses are intended for collection from a point light source inside the SiC, the focal length, $f$ is set to be 10 μm. Such a small focal length $f$ corresponds to a large numerical aperture. Therefore, the maximum diameter of the metalens is limited to 30μm, to allow for sufficient phase sampling at the outer Fresnel zones of the lens.

The second unconventional metalens design adds a cubic term to the parabolic phase, called a cubic metalens. The phase profile of such a lens is shown in Figure 1f and is described by the equation

$$\varphi_c(x,y) = \frac{2\pi}{\lambda}\left(f - \sqrt{x^2 + y^2 + f^2}\right) + \frac{\alpha}{L^3}(x^3 + y^3).$$

Here $\alpha$ is the extent of the cubic phase, and $L$ is the radius of the lens aperture. For our design, the cubic phase extent is $\alpha = 55\pi$ and determines the extent to which the focal spot is stretched along the propagation direction. This design can be adapted to the imaging of broadband spectrum, correcting the chromatic aberration caused by different wavelengths [31]. However, stretching the focal spot will also disperse energy that converges into the spot, thereby darkening the image without post-computational reconstruction.

Next, we fabricated both types of metalenses using a high-quality epitaxially grown layer of 4H-SiC. The lenses are patterned using electron beam lithography. Following the development of the resist, a metal mask is deposited on top, consisting of 10 nm/150 nm titanium/nickel. After liftoff, the sample was etched using an inductively coupled plasma reactive ion etching system (500 ICP, 15 RF, 3 sccm Ar, 3 sccm $O_2$, and 15 sccm of $SF_6$, 10 mTorr) until a depth of $H$=1,200 nm is achieved.

Figure 2a shows the scanning electron microscope (SEM) images of both types of metalenses (parabolic and cubic), featuring the parabolic metalens design on the left and the cubic metalens design on the right. Both metalenses are 30 microns in diameter, as per our designs. The apparent color gradient is an effect of the pillar density and size within the region. Figures 2b,c show close-up images of the regions defined in Figure 2a, where the larger pillars possess well-defined borders and separation from the neighboring ones. The smallest pillars are full-sized and thus will impart the necessary phase change on the lights that pass through.

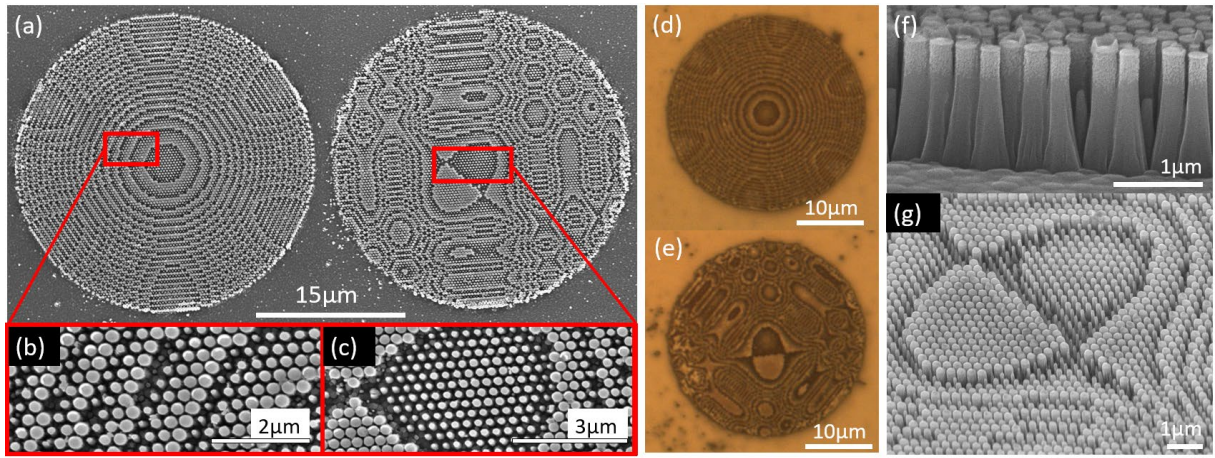

*Figure 2. Fabrication of SiC metasurfaces. (a) An SEM image of parabolic and cubic metalenses. (b, c) are zoomed-in regions of the respective metalens from (a), displaying the pillar top view. (d, e) Optical microscope images of the parabolic and cubic metalenses, respectively. (f, g) side view of the SiC pillars that form the metalenses.*

Figure 2d, e shows the optical images of the parabolic and cubic metalenses, respectively. Figures 2f,g further show an angled high-resolution SEM image of the side profile of the pillars. The pillars exhibit a small degree of tapering, which is in principle an undesired effect that could be improved in the future. Nevertheless, the fabricated SiC nanostructures fall within the layout margins of our design and allow for the proper functionality of the metalenses.

We now turn to the characterization of the optical properties of the metalenses. We employed a lab-built transmission setup, as shown in Figure 3a to characterize the focusing performance of the metalenses. A collimated laser beam (980 nm) passed through the metalens, and the light is collected using a high numerical aperture objective (NA=0.95) onto a CCD camera. The operation of the metalens is also tested by shining white light onto it and observing the focal point, as shown in Figure 3b. Despite not using the exact design wavelength, the focusing of the metalenses is clearly seen. The small wavelength deviation of the illumination laser will, however, result in a decrease in focussing efficiency with respect to the initial design.

To quantify the light focusing by the two types of metalenses, their focal spot was analyzed by capturing 100 images taken at 0.2 µm increments from the top surface of the metalens ($z = 0$). The images for each step are averaged and the horizontal transverse cross-sections as a function of the propagation distance are shown in Figures 3c,f, for the parabolic and the cubic lenses, respectively. Additionally, Figures 3d,g show the focal spots of the parabolic and cubic designed metalens in the *x-y* plane for the distances marked with a white dashed line in Figures 3c,f. For the parabolic metalens, the cross-section was taken at 3.5 µm above the surface of the metalens, Figure 3d. It shows a circular focal spot of a full width at half-maximum of FWHM=0.58 µm. For the cubic metalens, the transverse cross-section, taken at the same longitudinal distance of 3.5 µm, also shows a circular focal spot. However, the focal spot is elongated and displays higher noise, as seen in the transverse cross-section in Figure 3g.

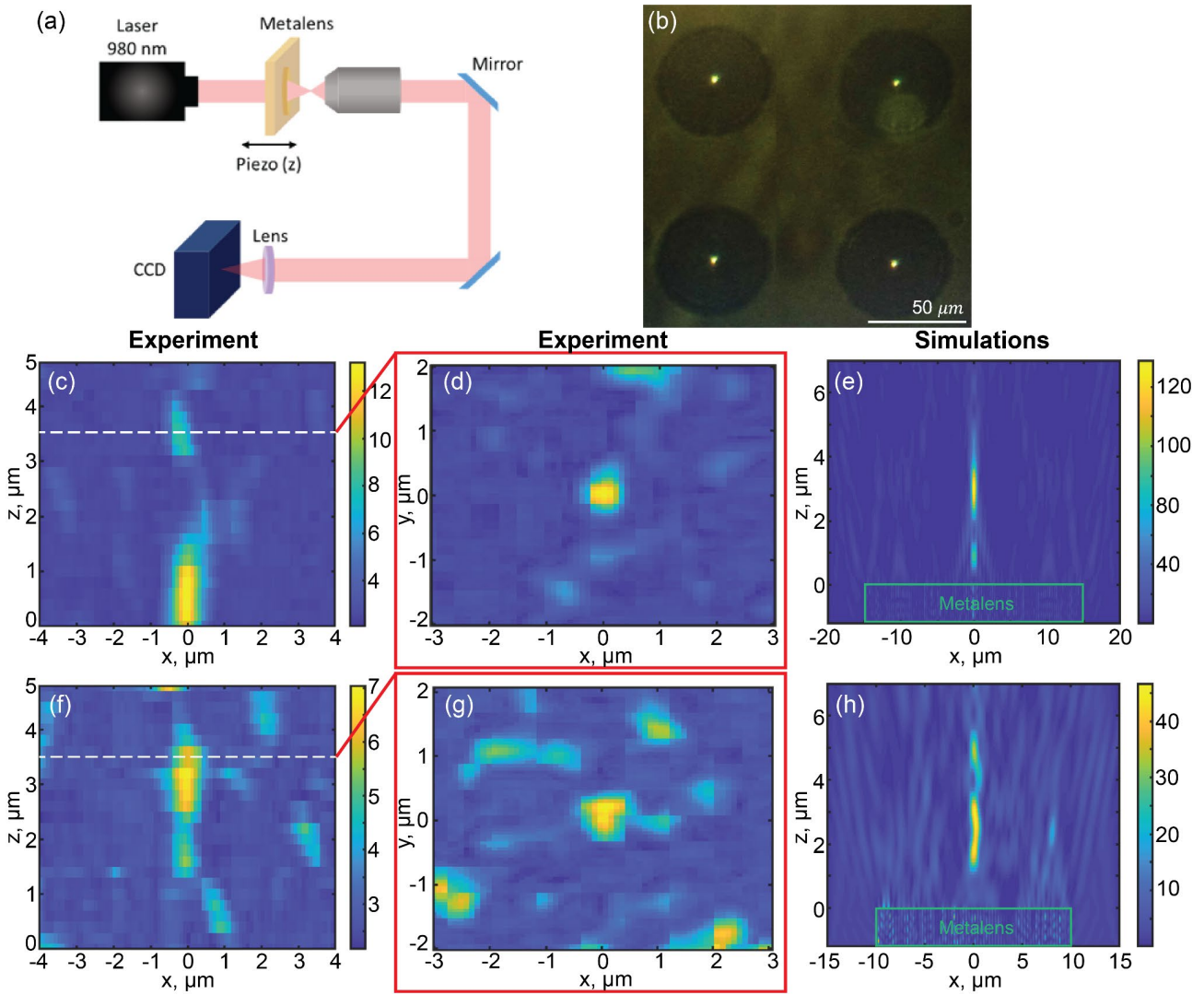

*Figure 3. Optical characterization of the metalenses. (a) Schematic of the measurement setup. (b) Optical images of the fabricated metalenses seen with the white light source of the optical microscope. The image shows the convergence focal spots after passing through the parabolic. (c) The light intensity from the illuminating 980 nm laser after the parabolic metalens, showing a bright spot from the surface of the metalens (at z = 0) up to approximately 2µm, with what appears to be designed focal spot at 3.5µm. (d) A cross-section from (c), taken at the designed focal distance of 3.5 µm. (e) Simulated data of the parabolic metalens, with a profile matching closely the measured data in (c). (f) The light intensity of the focal spot created by the cubic metalens, where the focal spot is extended from x = 2 µm to 4.5 µm. (g) Shows a cross-section from (f), taken at 3.5 µm where the spot in (e) is at the origin (0,0). (h) Simulated data of the cubic metalens, matching the observed extended focus observed in the experiments in the region 2-4 µm.*

To confirm our experiments and to provide a better insight on the metalens operation, we also performed numerical simulations of the light focusing after the metalens. The simulations are done using an FDTD method in Lumerical. We have imported the full metalens design into the software and have calculated the near-field light intensity with a 980nm plane wave illumination. Due to the asymmetric distribution of cubic metalens, the FDTD method cannot support such a large inhomogeneous structure without advanced symmetric boundary conditions. Therefore, in our simulations, we have reduced the diameter of the cubic metalens from 30 µm to 20 µm. The reconstructed focal points from the parabolic and cubic lenses are shown in Figure 3e,h, respectively. The reconstructed focal point for the parabolic and cubic metalenses confine the light in substantially different manners. Figure 3e depicts the focusing of the parabolic metalens, which shows the convergence of the input beam into a sharp focal spot. In this simulation, the FWHM of the focal spot taken at transverse cross-section at 3.5 µm is equal to 0.55. This value matches well the experimental measurements at the same longitudinal distance. However, the light intensity in the experiment is lower than in the simulation results, likely due to fabrication imperfections.

Figure 3h showcases the focusing of the cubic metalens. In this case, the geometry of metalens is designed to function chromatically, creating a broader focusing effect across the measured regions. The FWHM of the cubic metalens at longitudinal distance of 3.5 µm after the metalens is 0.67 µm, however, the transverse intensity profile shows a higher noise than the parabolic metalens simulation. Overall, the agreement with the experimental data is good, which assures the correct operation of the SiC metalenses for applications in SILs and other photonic applications.

To summarise, we have demonstrated the design, fabrication, and characterization of two distinct metalenses in monolithic silicon carbide. Both lenses show adequate performance and can be used to capture light from embedded light sources within silicon carbide. With further improvement and optimization of the fabrication process, and specifically eliminating the tapering and increasing the aspect ratio, better metalenses with improved focusing ability can be fabricated. Nevertheless, even in its current form, the metalenses are suitable for imaging embedded quantum emitters that operate at the near-infrared spectral range or are used to focus light from SiC LEDs.

**Acknowledgments**

We acknowledge the Australian Research Council (CE200100010) and the Asian Office of Aerospace Research and Development (FA2386-20-1-4014) for the financial support. The authors thank the ANFF (UTS node) for use of the nanofabrication facilities